\documentclass[reprint,preprintnumbers,prd,nofootinbib]{revtex4-1}

\usepackage{amsfonts}
\usepackage{amsmath}
\usepackage{array}
\usepackage{braket}
\usepackage{diagbox}
\usepackage{epstopdf}
\usepackage{graphicx}
\usepackage{hepparticles}
\usepackage{hepnicenames}
\usepackage{hepunits}
\usepackage{hyperref}
\usepackage[%
    utf8
]{inputenc}
\usepackage{multirow}
\usepackage{slashed}
\usepackage{subfigure}
\usepackage{placeins}
\usepackage[%
    normalem
]{ulem}
\usepackage[%
    usenames,
    svgnames,
    dvipsnames
]{xcolor}
\usepackage{xspace}

\newcommand{\order}[1]{\mathcal{O}\left({#1}\right)}

\newcommand{\reffig}[1]{figure~\ref{fig:#1}}

\newcommand{\reftab}[1]{table~\ref{tab:#1}}

\renewcommand{\theta}{\vartheta}

\newcommand{\para}{\parallel}

\newcommand{\EOS}{\texttt{EOS}\xspace}

\newcommand{\wilson}[1]{\mathcal{C}_{#1}}
\newcommand{\op}[1]{\mathcal{O}_{#1}}

\begin{document}

\allowdisplaybreaks

\preprint{EOS-2019-03, INT-PUB-19-042, TUM-HEP 1242/19}
\title{Bayesian analysis of $b\to s\mu^+\mu^-$ Wilson coefficients using the full angular distribution of $\Lambda_b\to \Lambda(\to p\, \pi^-)\mu^+\mu^-$ decays}

\author{Thomas Blake}
\email{thomas.blake@cern.ch}
\affiliation{Department of Physics, University of Warwick, Coventry CV4 7AL, UK}
\author{Stefan Meinel}
\email{smeinel@email.arizona.edu}
\affiliation{Department of Physics, University of Arizona, Tucson, AZ 85721, USA}
\author{Danny van Dyk}
\email{danny.van.dyk@gmail.com}
\affiliation{Physik Department, Technische Universit\"at M\"unchen, James-Franck-Stra\ss{}e 1, 85748 Garching b. M\"unchen, Germany}

\begin{abstract}
    Following updated and extended measurements of the full angular distribution of the decay
    $\Lambda_b\to \Lambda(\to p\,\pi^-)\mu^+\mu^-$ by the LHCb collaborations, as well as a
    new measurement of the $\Lambda \to p \pi^-$ decay asymmetry parameter by the BESIII collaboration,
    we study the impact of these results on searches for non-standard effects in exclusive $b\to s\mu^+\mu^-$ decays.
    To this end, we constrain the Wilson coefficients $\wilson{9}$ and
    $\wilson{10}$ of the numerically leading dimension-six operators in the weak effective Hamiltonian,
    in addition to the relevant nuisance parameters.
    In stark contrast to previous analyses of this decay mode, the changes in the updated experimental
    results lead us to find very good compatibility with both the Standard Model and with
    the $b\to s\mu^+\mu^-$ anomalies observed in rare $B$-meson decays.
    We provide a detailed analysis of the impact
    of the partial angular distribution, the full angular distribution,
    and the $\Lambda_b\to \Lambda\mu^+\mu^-$ branching fraction on the Wilson coefficients.
    In this process, we are also able to constrain the size of the production polarization
    of the $\Lambda_b$ baryon at LHCb.
\end{abstract}

\maketitle

\section{Introduction}
\label{sec:intro}

The persistent anomalies in the rare flavor-changing decays of $B$ mesons, which arise in analyses of branching fractions, angular distributions 
and lepton flavour universality tests, have sparked considerable
interest in constructing candidate theories to replace the Standard Model (SM) of particle physics; see for example
ref.~\cite{Buttazzo:2017ixm} for a comprehensive guide.
If these anomalies are indeed a hint of physics Beyond the SM (BSM), then we should see signs of similar
deviations in the baryonic partners of these rare $B$ meson decays, \emph{e.g.} in  $\Lambda_b\to\Lambda(\to p \pi^-)\mu^+\mu^-$.

The decay mode $\Lambda_b\to\Lambda(\to p \pi^-)\mu^+\mu^-$ is quite appealing from a theoretical point of view. Like the $B\to K^*(\to K \pi)\mu^+\mu^-$ decay,
it provides a large number of angular observables and is sensitive to all Dirac structures in the effective weak Hamiltonian \cite{Boer:2014kda,Blake:2017une,Das:2018sms,Yan:2019tgn}.
At the same time, because the $\Lambda$ baryon is stable under the strong interactions, lattice QCD calculations of the $\Lambda_b \to \Lambda$ form factors \cite{Detmold:2016pkz}
do not require a complicated finite-volume treatment of multi-hadron states, as would be necessary for a rigorous calculation of $B\to K^*(\to K \pi)$ form factors
\cite{Briceno:2014uqa}\footnote{%
    The lattice determination of the $B\to K^*$ form factors in ref.~\cite{Horgan:2013hoa} and Light-Cone Sum Rule (LCSR) estimates in refs.~\cite{Straub:2015ica,Gubernari:2018wyi,Gao:2019lta}
    treat the $K^*$ as if it is stable, leading to systematic uncertainties that are difficult to quantify; see ref.~\cite{Descotes-Genon:2019bud} for a first
    study of the finite width effects in LCSRs.
}.

A previous analysis of the constraints of $\Lambda_b\to\Lambda(\to p \pi^-)\mu^+\mu^-$ on the $b\to s\mu^+\mu^-$ Wilson coefficients \cite{Meinel:2016grj} using --- by now --- outdated experimental inputs found a central value of $\wilson{9}$ shifted in the opposite direction from the SM point compared to the $B$-meson findings.
In this paper we confront this previous analysis with new, updated, and reinterpreted experimental results,
and constrain BSM effects in $b\to s\mu^+\mu^-$ operators.

\section{Framework}
\label{sec:framework}

We use the standard weak effective field theory that describes
flavour-changing neutral $b\to s \lbrace \mu^+\mu^-, \gamma, q\bar{q}\rbrace$ transitions
up to mass-dimension six~\cite{Buchalla:1995vs}.
Following the conventions in ref.~\cite{Bobeth:2012vn}, the effective Hamiltonian can be expressed as
\begin{equation}
\begin{aligned}
  \label{eq:Heff}
  {\cal{H}}_{\rm eff}
  & = - \frac{4\, G_F}{\sqrt{2}}  V_{tb}^{} V_{ts}^* \,\frac{\alpha_e}{4 \pi}\,
       \sum_i \wilson{i}(\mu)  \op{i}\\
  & + \order{V_{ub} V_{us}^*} + \text{h.c.}\,,
\end{aligned}
\end{equation}
where $G_F$ denotes the Fermi constant as extracted from muon decays, $V_{ij}$ are CKM matrix elements, and
$\alpha_e$ is the electromagnetic coupling at the scale of the $b$-quark mass, $m_b$. We write the short-distance (Wilson) coefficients as $\wilson{i}(\mu)$,
taken at a renormalization scale $\mu \simeq m_b$, and
 long-distance physics is expressed through matrix elements of the
effective field operators, $\op{i}$. For the decay in hand, the numerically
leading operators are
\begin{equation}
\label{eq:SM:ops}
\begin{aligned}
    \op{7(7')} & = \frac{m_b}{e}\!\left[\bar{s} \sigma^{\mu\nu} P_{R(L)} b\right] F_{\mu\nu}\,,\\
    \op{9(9')} & = \left[\bar{s} \gamma_\mu P_{L(R)} b\right]\!\left[\bar{\ell} \gamma^\mu \ell\right]\,,\\
    \op{10(10')} & = \left[\bar{s} \gamma_\mu P_{L(R)} b\right]\!\left[\bar{\ell} \gamma^\mu \gamma_5 \ell\right]\,.
\end{aligned}
\end{equation}
A prime indicates a flip of the quarks' chiralities with respect to the
unprimed, Standard Model(SM)-like operator.
The ten form factors describing the hadronic matrix elements $\langle \Lambda |\bar{s}\Gamma b|\Lambda_b\rangle$ for $\Gamma\in\{\gamma_\mu,\gamma_\mu\gamma_5,\sigma_{\mu\nu}  \}$ are taken from the lattice QCD calculation of ref.~\cite{Detmold:2016pkz}.
The inclusion of non-local charm effects
follows the usual Operator Product Expansion (OPE) at large momentum transfer $q^2$ in combination with the assumption of global quark-hadron duality; see
refs.~\cite{Grinstein:2004vb,Beylich:2011aq} for the theoretical basis and
ref.~\cite{Boer:2014kda} for the phenomenological application to $\Lambda_b\to \Lambda\mu^+\mu^-$ decays.
At leading power in the OPE, the matrix elements can be expressed in terms of the aforementioned form factors.
The uncertainty of the form factors and the breaking of the quark-hadron duality assumption are treated through
a large set of nuisance parameters in the same way as discussed in ref.~\cite{Meinel:2016grj}.
\\

\begin{table}[t]
\begin{center}
\renewcommand{\arraystretch}{1.4}
\begin{tabular}{l ccc}
\toprule
  Quantity & Prior & Unit & Reference\\
\colrule
  \multicolumn{4}{c}{CKM Wolfenstein parameters}\\
\colrule
  $A$                            &  $0.826 \pm 0.012$      &  ---      &  \cite{Bona:2006ah}\\
  $\lambda$                      &  $0.225 \pm 0.001$      &  ---      &  \cite{Bona:2006ah}\\
  $\bar{\rho}$                   &  $0.148 \pm 0.043$      &  ---      &  \cite{Bona:2006ah}\\
  $\bar{\eta}$                   &  $0.348 \pm 0.010$      &  ---      &  \cite{Bona:2006ah}\\
\colrule
  \multicolumn{4}{c}{$B_s$ decay constant}
\\
\colrule
  $f_{B_s}$                      &  $230.7 \pm 1.3$        &  $\MeV$   &  \cite{Bazavov:2017lyh}\\
\colrule
  \multicolumn{4}{c}{$\Lambda\to p\,\pi^-$ decay parameter}
\\
\colrule
  $\alpha$                      &  $0.750 \pm 0.010$       &  ---      &  \cite{Ablikim:2019vaj} \\
\colrule
  \multicolumn{4}{c}{duality violation in the $\Lambda_b\to \Lambda\mu^+\mu^-$ amplitudes}
\\
\colrule
  $r_{i,J_z}$, $i=\perp,\para$, $J_z = 0,1$
                                &  $0.0 \pm 0.03$          &  ---      &  \cite{Meinel:2016grj} \\
\botrule
\end{tabular}
\renewcommand{\arraystretch}{1.0}
\caption{%
    Prior distributions of selected nuisance parameters: the Cabibbo-Kobayashi-Maskawa (CKM) parameters,
    the decay constant of the $B_s$, $f_{B_s}$, and the $\Lambda\to p\,\pi^-$ parity-violating decay parameter, $\alpha$.
    For the CKM parameters, we use the Summer'18 update of a Bayesian analysis
    of only tree-level decays performed by the UTfit Collaboration \cite{Bona:2006ah}.
    All distributions are Gaussian.
    The prior distribution for the $\Lambda_b\to \Lambda$ form
    factors is a multivariate Gaussian with inputs directly taken from the
    lattice QCD calculation in ref.~\cite{Detmold:2016pkz}.
}
\label{tab:nuisance-parameters}
\end{center}
\end{table}

We define four fit scenarios labeled ``SM($\nu$-only)'', $(9)$, ``$(9,10)$'' and ``$(9,10,9',10')$'':
\begin{equation}
\label{eq:scenarios}
\begin{aligned}
    \text{SM($\nu$-only)} & :
    \begin{cases}
        \wilson{9,10}     & \text{SM values}\\
        \wilson{9',10'}   & \text{SM values}\\
        \vec\nu           & \text{within priors}
    \end{cases}\,,\\
    (9) & :
    \begin{cases}
        \wilson{9}        & \in [-1,+9]\\
        \wilson{9,9',10'} & \text{SM values}\\
        \vec\nu           & \text{within priors}
    \end{cases}\,,\\
    (9,10) & :
    \begin{cases}
        \wilson{9}        & \in [-1,+9]\\
        \wilson{10}       & \in [-9,-1]\\
        \wilson{9',10'}   & \text{SM values}\\
        \vec\nu           & \text{within priors}
    \end{cases}\,,\\
    (9,10,9',10') & :
    \begin{cases}
        \wilson{9}        & \in [-1,+9]\\
        \wilson{10}       & \in [-9,-1]\\
        \wilson{9',10'}   & \in [-10,+10]\\
        \vec\nu           & \text{within priors}
    \end{cases}\,.
\end{aligned}
\end{equation}
In the above, $\vec\theta = (\wilson{9})$, $\vec\theta = (\wilson{9}, \wilson{10})$
or $\vec\theta = (\wilson{9}, \wilson{10}, \wilson{9'}, \wilson{10'})$
denotes the parameters of interest. Nuisance parameters $\vec{\nu}$
emerge in the parametrization of the (local) hadronic matrix elements
in terms of $\Lambda_b\to \Lambda$ form factors; in the amount of parity violation
in $\Lambda\to p\,\pi^-$ decays ($\alpha_{\Lambda \to p\,\pi^-}$);
and when accounting for duality violating effects that go beyond the low-recoil OPE.
The values of the nuisance parameters are given in \reftab{nuisance-parameters}.
Our statistical setup is identical to the one in \cite{Meinel:2016grj}.

\section{Data}

\begin{figure}[t]
    \includegraphics[width=.49\textwidth]{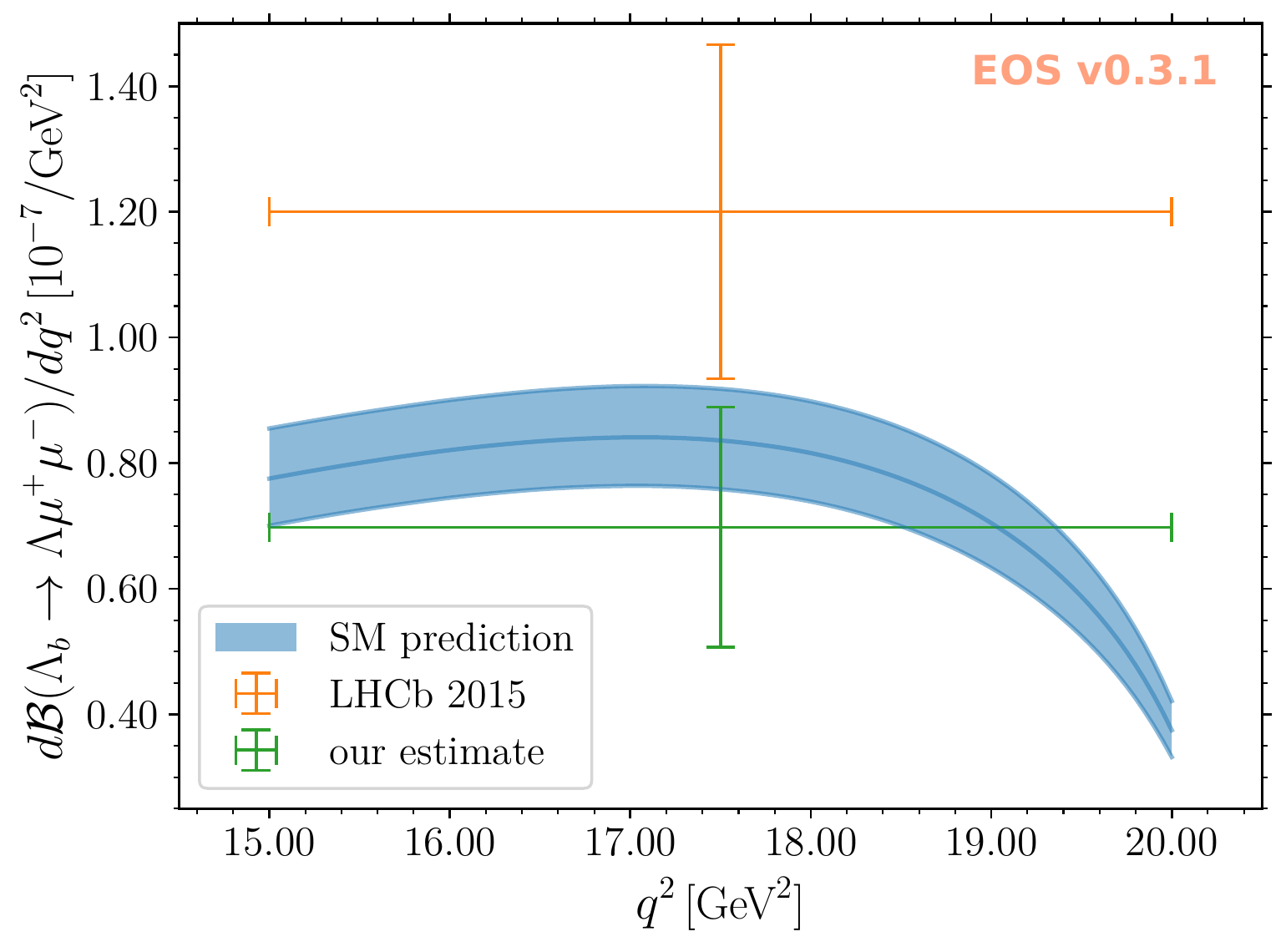}
    \caption{
        Comparison of the predicted differential branching fraction for the $\Lambda_b\to\Lambda\mu^+\mu^-$ decay 
        in the SM with the measured result by LHCb in the bin $15\,\GeV^2 \leq q^2\leq 20\,\GeV^2$,
        alongside our reinterpreted result. The central curve and band show the median and
        $68\%$ probability envelope of the prior predictions of the branching fraction.
        Note that while the OPE prediction cannot reproduce the resonant structures arising in the differential
        distributions, it is expected to reasonably describe the charm effects in $q^2$-integrated observables,
        up to small duality-violating effects.
    }
    \label{fig:Lb-to-Lmumu-BR}
\end{figure}

The following new experimental results supersede those used in the previous analysis in ref.~\cite{Meinel:2016grj}:
\begin{enumerate}
    \item The BESIII collaboration has recently measured~\cite{Ablikim:2019vaj}
    the parity-violating parameter $\alpha$ in $\Lambda \to p\,\pi^-$ decays in
    $e^+e^-\to J/\!\psi \to \Lambda\bar\Lambda$ production.
    This measurement is incompatible with the previous world average from secondary scattering data~\cite{Tanabashi:2018oca}.
    Given the inability to validate
    assumptions and intermediate results used in the measurements entering the previous
    world average of $\alpha$, the Particle Data Group (PDG) has replaced their previous average
    with the BESIII measurement for the upcoming ``Review of Particle Physics''.
    We use the new BESIII result in this paper.

    \item The LHCb collaboration has recently published~\cite{Aaij:2018gwm} their
    measurement of the complete set of angular observables in decays of
    polarised $\Lambda_b$ baryons to $\Lambda\mu^+\mu^-$ final states. This
    supersedes the three angular observables measured in ref.~\cite{Aaij:2015xza}.  Of
    particular interest is an erratum to the 2015 LHCb measurement~\cite{Aaij:2015xza}, which
    explains that the reported result for the leptonic forward-backward
    asymmetry $A^\ell_{\text{FB}}$ was misattributed. In effect LHCb had accidentally reported the value of the $C\!P$-asymmetry of this
    observable, rather than its $C\!P$-average.

    \item The ATLAS, CMS, and LHCb collaborations have 
    each measured \cite{Aaboud:2018mst,CMS:2019qnb,Aaij:2017vad} the
    time-integrated branching ratio of the decay $B_s \to \mu^+\mu^-$, denoted
    here as $\overline{\mathcal{B}}(\bar{B}_s\to \mu^+\mu^-)$ \cite{DeBruyn:2012wk}.
    Within our fit scenarios, the combination $|C_{10}-C_{10'}|$ is constrained by these measurements.

    \item The LHCb measurement of the $\Lambda_b\to \Lambda\mu^+\mu^-$ branching fraction is normalized to the $\Lambda_b\to \Lambda J/\psi$ fraction.
    In converting this relative ratio to an absolute branching fraction, LHCb used the PDG world average for the product~\cite{Aaij:2015xza}
    \begin{equation*}
        f(b\to \Lambda_b) \times \mathcal{B}(\Lambda_b\to \Lambda J/\!\psi)\,,
    \end{equation*}
    where $f(b\to \Lambda_b)$ is the $\Lambda_b$ fragmentation fraction. The LHCb measurement used an old average of $f(b\to \Lambda_b)$ that included measurements from the LEP and TeVatron experiments.
    The fragmentation fraction as a function of the $b$-quark transverse momentum has since been measured by the LHCb collaboration \cite{Aaij:2014jyk}.
    Given the strong dependence on the $b$-quark production processes and the $b$-quark transverse momentum, combining the LEP and TeVatron results appears unwise. 
    Hence, we remove the LEP results from the average, and
    calculate the branching fraction of the $\Lambda_b\to \Lambda\mu^+\mu^-$ decay anew, using only the average of the TeVatron results.
    This calculation follows the approach by the Heavy Flavour Averaging group in Ref.~\cite{Amhis:2016xyh}.
    The $\Lambda_b$ production fraction is derived from  $f(b \to {\rm baryon}) =  0.218 \pm 0.047$, assuming isospin symmetry in $\Xi_{b}^{0}$ and $\Xi_{b}^{-}$ production, \emph{i.e.}
    \begin{align}
    \begin{split}
    f(b & \to {\rm baryon})= \\ & f(b \to {\Lambda_b}) + 2 f(b\to\Xi_{b}^{-}) + f(b\to{\Omega_b^-})~.
    \end{split}
    \end{align}
    An updated value for $f(b\to\Lambda_b)$ is determined using the ratios $f(b \to {\Xi_b^-})/f(b\to {\Lambda_b})$ and $f(b\to{\Omega_b^-})/f(b\to {\Lambda_b})$ from ref.~\cite{PDG2018}, assuming equal partial widths for the $\Lambda_b \to J/\!\psi \Lambda$, $\Xi_b^{-} \to J/\!\psi \Xi^{-}$ and $\Omega_{b}^{-} \to J/\!\psi \Omega^{-}$ decays. 
    The updated value of $f(b\to\Lambda_b)$ results in an updated branching fraction for the $\Lambda_b \to J/\!\psi \Lambda$ decay of ${\cal B}(\Lambda_b \to J/\!\psi \Lambda) = (3.7\pm 1.0 )\times 10^{-4}$. 
    Using this branching fraction value we obtain, for the bin $15\,\GeV^2 \leq q^2 \leq 20\,\GeV^2$,\linebreak
    \begin{align}
    \begin{split}
        \mathcal{B}(\Lambda_b\to & \Lambda\mu^+\mu^-)_{[15,20]}
            = \\ & (3.49 \pm 0.26 \pm 0.92) \times 10^{-7}\,.
    \end{split}
    \end{align}
    This is significantly smaller than the branching fraction reported by LHCb in ref.~\cite{Aaij:2015xza}.
    This result, alongside the original, unmodified, LHCb result for the branching ratio and the SM predictions
    for the differential branching ratio is juxtaposed in \reffig{Lb-to-Lmumu-BR}.

    \item The fits of ref.~\cite{Meinel:2016grj} include data on the inclusive $B\to X_s \ell^+\ell^-$ branching fraction.
    Given the improved precision of the $\Lambda_b\to \Lambda\mu^+\mu^-$ results and the $\bar{B}_s\to \mu^+\mu^-$ branching fraction, this is no longer necessary.
\end{enumerate}

For the following fits we define three data sets entering the likelihood:
\begin{description}
    \item[data set 1] includes the three measurements of $\mathcal{B}(\bar{B}_s\to \mu^+\mu^-)$ and the LHCb measurement of the nine independent angular observables in the $\Lambda_b\to \Lambda(\to p\, \pi^-)\mu^+\mu^-$ angular distribution for an unpolarized $\Lambda_b$ baryon;

    \item[data set 2] includes the three measurements of $\mathcal{B}(\bar{B}_s\to \mu^+\mu^-)$ and the he LHCb measurement of the 33 independent angular observables in the $\Lambda_b\to \Lambda(\to p\, \pi^-)\mu^+\mu^-$ angular distribution for a polarized $\Lambda_b$ baryon;

    \item[data set 3] contains data set 2, but also includes the reinterpreted branching ratio of $\Lambda_b\to \Lambda \mu^+\mu^-$
        decays.
\end{description}
Our nominal data set, which we use for our main results and conclusions, is \emph{data set 2}.

\section{Results}
\label{sec:results}

\begin{figure}[b]
\includegraphics[width=.40\textwidth]{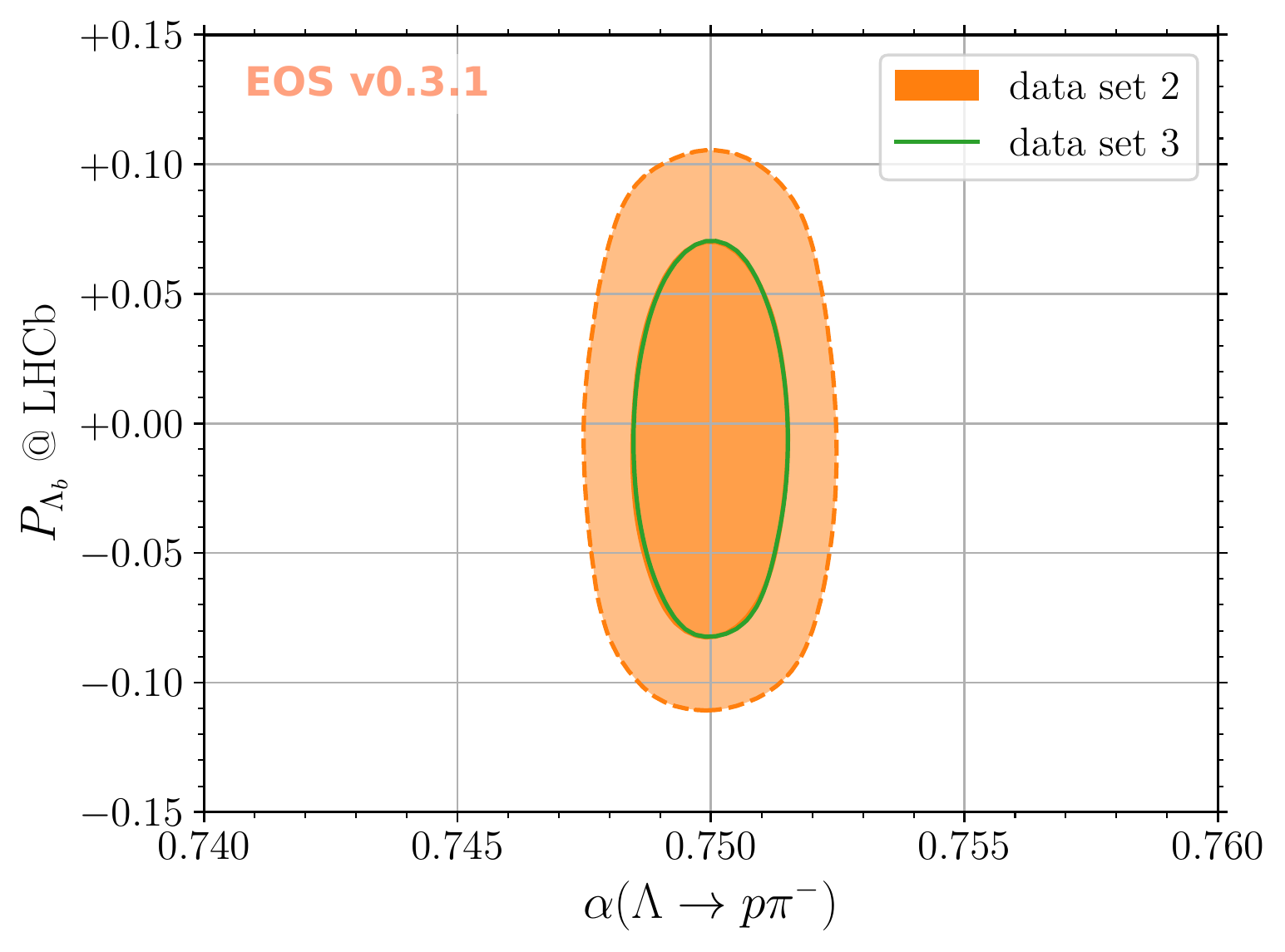}
\caption{
Contours for the joint 2D posterior for the asymmetry parameter, $\alpha$, and the $\Lambda_b$ production polarisation at LHCb, $P_{\Lambda_b}$.
We show $68\%$ probability contours for data sets 2 and 3, and the 95\% contour for data set 2.
}
\label{fig:alphaP}
\end{figure}

\begin{table*}[t!]
    \renewcommand{\arraystretch}{1.25}
    \resizebox{\textwidth}{!}{%
    \begin{tabular}{l c c c c c c c c c c c c c}
        \toprule
        ~                                      &
            ~                                  &
            \multicolumn{3}{c}{SM($\nu$-only)} &
            \multicolumn{3}{c}{$(9)$}          &
            \multicolumn{3}{c}{$(9,10)$}       &
            \multicolumn{3}{c}{$(9,10,9',10')$}\\
        Contribution                           &
            \diagbox{\footnotesize \#obs}{\footnotesize \#par}             &
            $0$                                & 
            $0$                                & 
            $0$                                & 
            $1$                                & 
            $1$                                & 
            $1$                                & 
            $2$                                & 
            $2$                                & 
            $2$                                & 
            $4$                                & 
            $4$                                & 
            $4$                                \\
        \colrule
        $\mathcal{B}(\bar{B}_s\to\mu^+\mu^-)$  &
            $3$                                &
            $ 1.87$                            & 
            $ 1.87$                            & 
            $ 1.84$                            & 
            $ 1.87$                            & 
            $ 1.87$                            & 
            $ 1.83$                            & 
            $ 0.04$                            & 
            $ 0.04$                            & 
            $ 0.04$                            & 
            $ 0.04$                            & 
            $ 0.04$                            & 
            $ 0.11$                            \\
        ang.~obs. (unpol.)                     &
            $9$                                &
            $ 7.85$                            & 
            ---                                &
            ---                                &
            $ 7.60$                            & 
            ---                                &
            ---                                &
            $ 7.68$                            & 
            ---                                &
            ---                                &
            $ 7.39$                            & 
            ---                                &
            ---                                \\
        ang.~obs. (all)                        &
            $33$                               &
            ---                                &
            $43.11$                            & 
            $43.11$                            & 
            ---                                &
            $42.72$                            & 
            $42.82$                            & 
            ---                                &
            $42.83$                            & 
            $42.84$                            & 
            ---                                &
            $42.28$                            & 
            $42.33$                            \\
        $\mathcal{B}(\Lambda_b\to\Lambda \mu^+\mu^-)$ &
            $1$                                &
            ---                                &
            ---                                &
            $ 0.06$                            & 
            ---                                &
            ---                                &
            $ 0.19$                            & 
            ---                                &
            ---                                &
            $ 0.00$                            & 
            ---                                &
            ---                                &
            $ 0.04$                            \\
        \colrule
        $\Lambda_b\to\Lambda$ form factors \hspace{-4ex}             &
            ---                              &
            $ 0.12$                          & 
            $ 0.28$                          & 
            $ 0.28$                          & 
            $ 0.10$                          & 
            $ 0.25$                          & 
            $ 0.29$                          & 
            $ 0.11$                          & 
            $ 0.26$                          & 
            $ 0.26$                          & 
            $ 0.07$                          & 
            $ 0.15$                          & 
            $ 0.16$                          \\
        \colrule
        \multirow{3}{*}{total}                 &
            $12$                               &
            $ 9.84$                            & 
            ---                                &
            ---                                &
            $ 9.57$                            & 
            ---                                &
            ---                                &
            $ 7.83$                            & 
            ---                                &
            ---                                &
            $ 7.49$                            & 
            ---                                &
            ---                                \\
                                               &
            $36$                               &
            ---                                &
            $45.25$                            & 
            ---                                &
            ---                                &
            $44.83$                            & 
            ---                                &
            ---                                &
            $43.14$                            & 
            ---                                &
            ---                                &
            $42.47$                            & 
            ---                                \\
                                               &
            $37$                               &
            ---                                &
            ---                                &
            $45.30$                            & 
            ---                                &
            ---                                &
            $45.13$                            & 
            ---                                &
            ---                                &
            $43.14$                            & 
            ---                                &
            ---                                &
            $42.58$                            \\
        \colrule
        $p$ value                              &
            ---                                &
            $ 0.63$                            & 
            $ 0.11$                            & 
            $ 0.13$                            & 
            $ 0.57$                            & 
            $ 0.10$                            & 
            $ 0.12$                            & 
            $ 0.65$                            & 
            $ 0.11$                            & 
            $ 0.14$                            & 
            $ 0.48$                            & 
            $ 0.08$                            & 
            $ 0.10$                            \\
        \colrule
        $\log_{10} \text{evidence}$            &
            ---                                &
            $ 22.41$                           & 
            $ 32.84$                           & 
            $ 39.41$                           & 
            $ 21.98$                           & 
            $ 32.36$                           & 
            $ 38.73$                           & 
            $ 21.28$                           & 
            $ 31.65$                           & 
            $ 38.11$                           & 
            $ 19.37$                           & 
            $ 29.87$                           & 
            $ 35.93$                           \\
        \botrule
    \end{tabular}
    }
    \renewcommand{\arraystretch}{1}
    \caption{%
        Summary of the goodness of fit for all combinations of fit models and data sets.
        We present the $\chi^2$ values for each contribution to the total likelihood, the $\chi^2$ of the total likelihood,
        and the corresponding $p$ value at the respective best-fit points. For the purpose of a Bayesian
        model comparison we also present the model evidence for each fit.
    }
    \label{tab:gof}
\end{table*}

\begin{figure}
    \includegraphics[width=.40\textwidth]{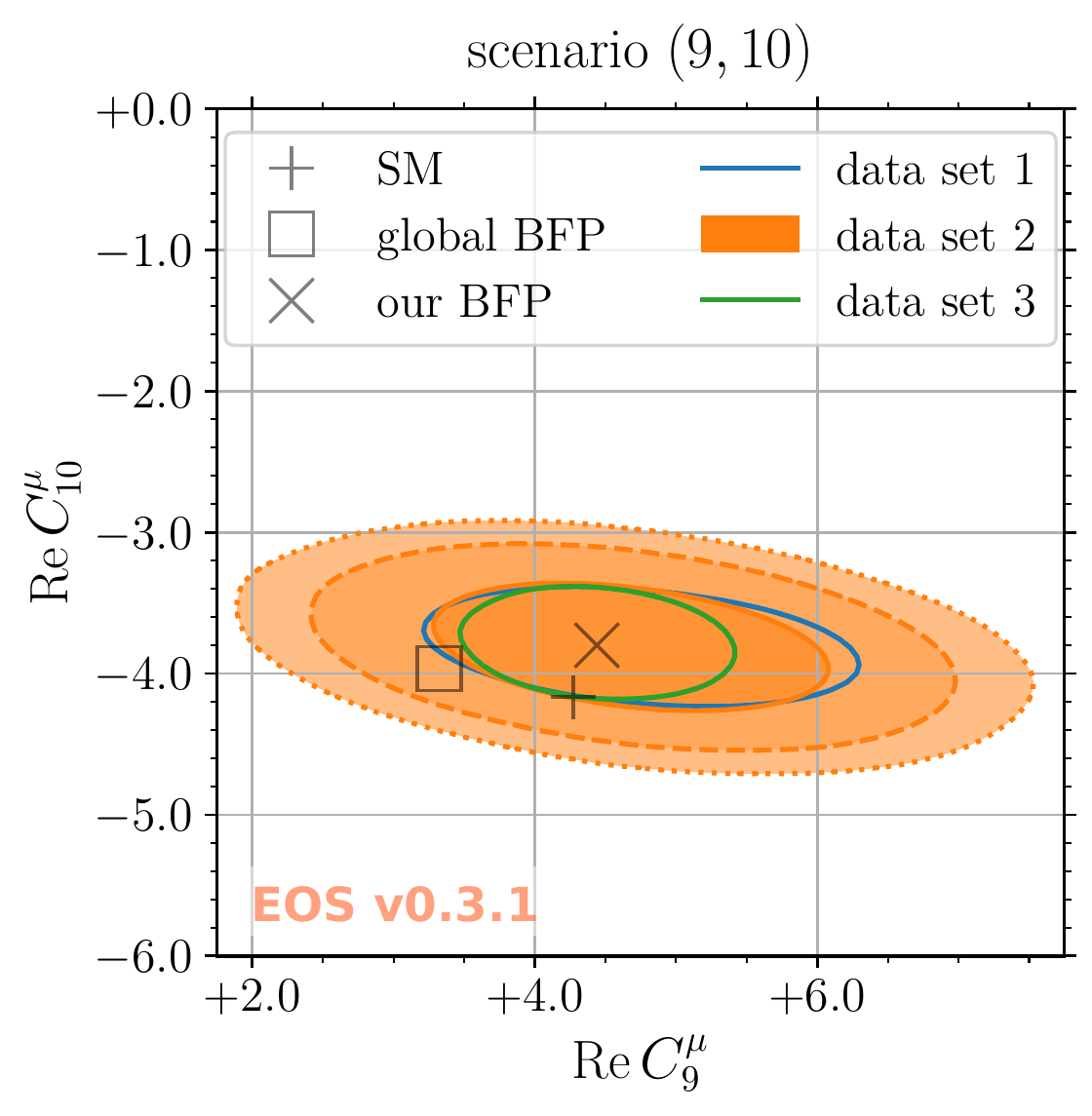}
    \caption{Contours of the joint 2D posterior for the parameters $\wilson{9}$ and $\wilson{10}$ in scenario $(9,10)$.
    All three data sets are used for both plots.
    We show $68\%$ probability contours for all
    data sets, and in addition $95\%$ and $99\%$ contours for our nominal data set 2.
    }
    \label{fig:c9c10}
\end{figure}

\begin{figure}
    \includegraphics[width=.40\textwidth]{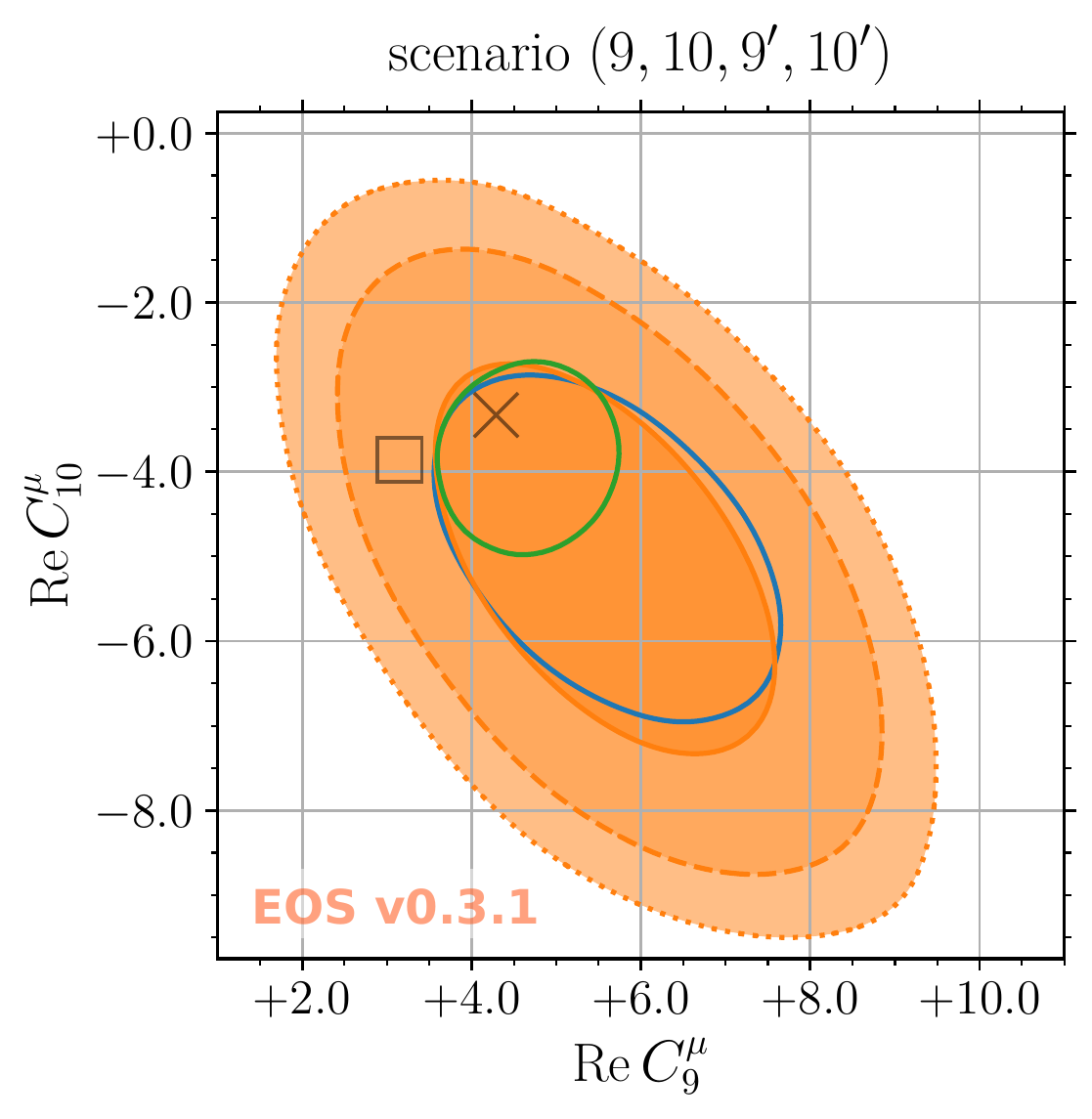}\\
    \includegraphics[width=.40\textwidth]{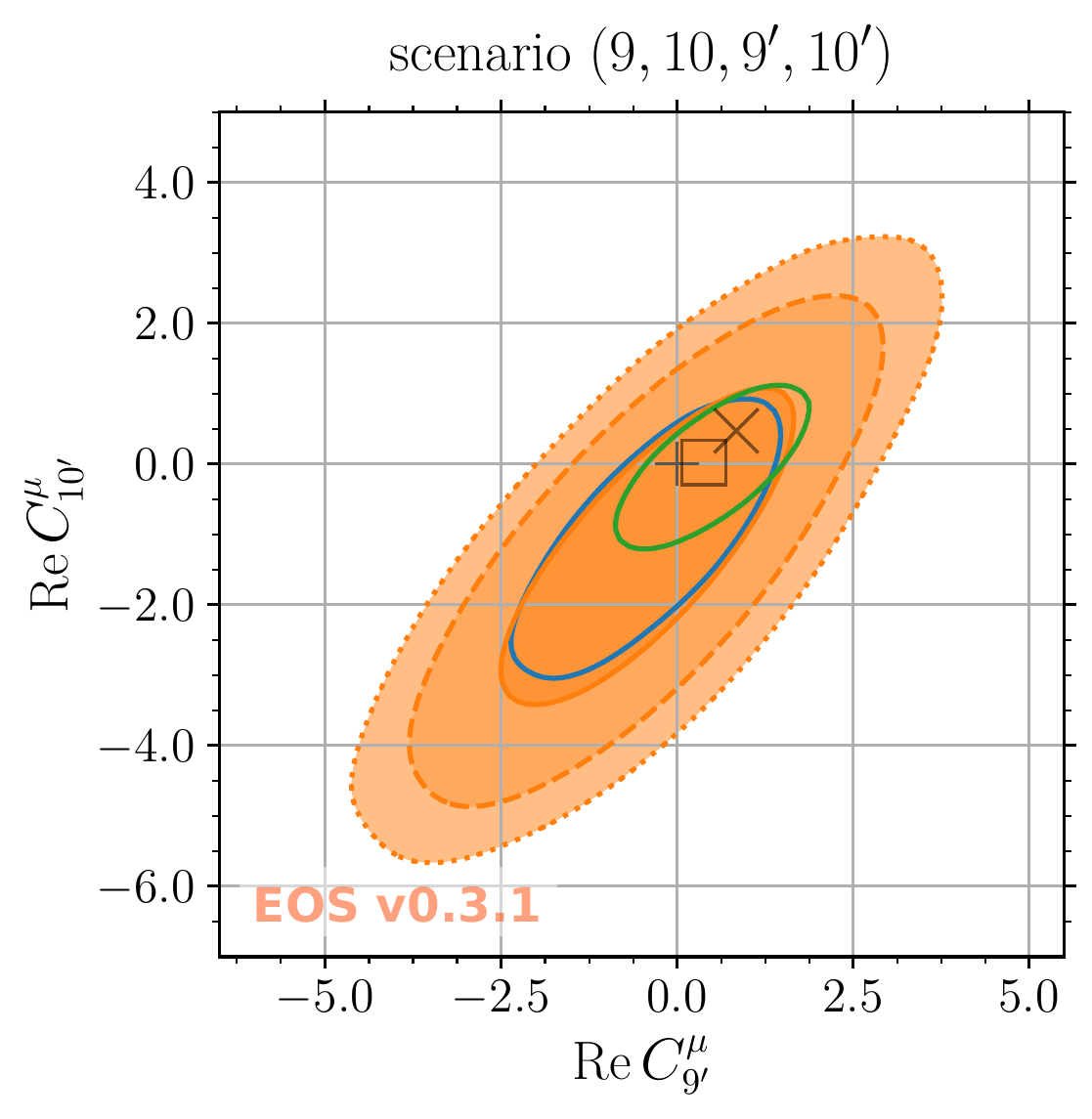}
    \caption{Contours of the joint 2D posterior for (top) the parameters $\wilson{9}$ and $\wilson{10}$
    and (bottom) the parameters $\wilson{9'}$ and $\wilson{10'}$ in scenario $(9,10,9',10')$.
    We show $68\%$ probability contours for all three
    data sets, and in addition $95\%$ and $99\%$ contours for our nominal data set 2.
    The legend is the same as in \reffig{c9c10}.
    }
    \label{fig:c99pc1010p}
\end{figure}

We use \EOS \cite{EOS} to carry out 12 fits for the three data sets and four fit scenarios. Summaries of the goodness of fit
in their respective best-fit points are collected in \reftab{gof}. Our findings are summarized as follows:
\begin{enumerate}
    \item The $\Lambda_b\to \Lambda(\to p\, \pi^-)\mu^+\mu^-$ angular distribution is compatible with the SM prediction, with acceptable $p$ values
    larger than $11\%$ for all three data sets.
    
    \item The $\Lambda_b$ polarization is compatible with zero in all four fit scenarios. We
    find $P_{\Lambda_b} = (0 \pm 5)\%$ at $68\%$ probability, and an upper limit for the magnitude of the polarization of $|P_{\Lambda_b}| \leq 11\%$ at
    $95\%$ probability (see fig.~\ref{fig:alphaP}); these results are independent of the choice of fit scenario.
    We show the two-dimensional marginalized posterior for the polarization and the decay parameter $\alpha$
    in \reffig{c9c10}.
    
    \item In the $(9)$ scenario, the $p$ values decrease slightly for all three data sets, with
    the minimal value of $10\%$ still acceptable. The best-fit point in our nominal fit using data set 2 is:
    \begin{equation}
    \begin{aligned}
        \wilson{9}  & = 4.8 \pm 0.8\,.
    \end{aligned}
    \end{equation}
    
    \item In the $(9,10)$ scenario, the $p$ values of all three data sets are slightly higher
    than in the SM. 
    The best-fit point in our nominal fit using  data set 2 is:
    \begin{equation*}
    \begin{aligned}
        \wilson{9}   & = +4.4 \pm 0.8\,, &
        \wilson{10}  & = -3.8 \pm 0.3\,.
    \end{aligned}
    \end{equation*}
    We find compatibility with the best-fit point
    obtained in rare semileptonic $B$ meson decays~\cite{Capdevila:2017bsm} at $\simeq 1.2\sigma$, and compatibility
    with the SM point at $\simeq 1 \sigma$. We show the two-dimensional marginalized posterior in \reffig{c9c10}.

    \item In the $(9,10,9',10')$ scenario, the $p$ values of all three data sets are lower
    than in the SM, with a minimal value of $8\%$.
    The best-fit point in our nominal fit using data set 2 is:
    \begin{equation*}
    \begin{aligned}
        \wilson{9}   & = +4.3 \pm 0.9\,, &
        \wilson{10}  & = -3.3 \pm 0.7\,, \\
        \wilson{9'}  & = +0.8 \pm 0.8\,, &
        \wilson{10'} & = +0.5 \pm 0.7\,.
    \end{aligned}
    \end{equation*}
    We find compatibility with the best-fit point obtained in rare semileptonic $B$ meson decays at $\simeq 1.5\sigma$,
    and compatibility with the SM point at less than $1\sigma$.  We show the two-dimensional marginalized posteriors in \reffig{c99pc1010p}.
    
    \item We compute the model evidence for all combinations of data sets and fit scenarios. Our
    results are listed in \reftab{gof}. From these results we compute the Bayes factors:
    \begin{equation*}
    \begin{aligned}
        \log_{10} \frac{P(\text{data set 2}\,|\,(9))          }{P(\text{data set 2}\,|\,\text{SM($\nu$-only)})} & = -0.48\,, \\
        \log_{10} \frac{P(\text{data set 2}\,|\,(9,10))       }{P(\text{data set 2}\,|\,\text{SM($\nu$-only)})} & = -1.15\,, \\
        \log_{10} \frac{P(\text{data set 2}\,|\,(9,10,9',10'))}{P(\text{data set 2}\,|\,\text{SM($\nu$-only)})} & = -2.97\,. \\
    \end{aligned}
    \end{equation*}
    According to Jeffrey's interpretation of the Bayes factor \cite{Jeffreys:1961}, we find
    the degree to which the scenario SM($\nu$-only) is favoured over scenarios $(9)$, $(9,10)$, and $(9,10,9',10')$
    to be \emph{barely worth mentioning}, \emph{strong}, and \emph{decisive}, respectively.
\end{enumerate}

\section{Conclusion}
\label{sec:summary}

We carry out the first Beyond the Standard Model (BSM) analysis of the measurements of the full angular distribution in $\Lambda_b\to\Lambda(\to p\pi)\mu^+\mu^-$ decays.
In this analysis we challenge the available data in four fit scenarios, corresponding to the absence of BSM effects (scenario SM($\nu$-only)); BSM effects only
in operators present in the SM (scenarios $(9)$ and $(9,10)$); and BSM effects in all (axial)vector operators (scenario $(9,10,9',10')$).
Our results supersede those of a previous analysis of this decay mode in ref.~\cite{Meinel:2016grj}, due to updates to various experimental results and a correction in the numerical code. 

The best-fit points in our three BSM scenarios are compatible with both the SM and the best-fit points obtained from phenomenological analyses of
exclusive $b\to s\mu^+\mu^-$ decays of $B$ mesons. The overall compatibility between such fits to the rare $\Lambda_b$ decay observables and the rare $B$ decay observables
has significantly improved since the previous analysis~\cite{Meinel:2016grj}.
The primary reason for this improvement is the use of an entirely new data
set that corrects an error in the measurement of the leptonic forward-backward
asymmetry. Another change is the removal of the inclusive $B\to X_s\ell^+\ell^-$ branching fractions from the fit.
For data set 3, we also use an updated value for the $\Lambda_b \to \Lambda \mu^+\mu^-$ branching ratio that is substantially smaller than
what was used in the previous analysis. Finally, we corrected an error in the handling of the tensor
form factors within \EOS (fixed as of v0.3~\cite{EOS}), which reduces the predicted branching fraction by a small amount
and affects the BSM interpretation.

We find that the scenarios SM($\nu$-only) and $(9)$ are almost equal in their efficiency of describing the $\Lambda_b\to \Lambda\mu^+\mu^-$ data. Moreover,
the remaining scenarios $(9,10)$ and $(9,10,9',10')$ are \emph{strongly} and \emph{decisively} disfavoured in a Bayesian model comparison.
\\

As a side result of our BSM analysis, we infer $P_{\Lambda_b}^\text{LHCb}$, the $\Lambda_b$ polarization in the LHCb phase space, from a rare decay for the first time.
We find $P_{\Lambda_b}^\text{LHCb} = (0\pm 5)\%$ at $68\%$ probability. This bound is independent of the fit scenarios, and is competitive with value obtained in the LHCb analysis
of $\Lambda_b\to \Lambda J/\!\psi$ decays of $(6\pm7\%)$~\cite{Aaij:2013oxa}.

\acknowledgments

We would like to thank David Straub for numerical comparisons of the $\Lambda_b\to\Lambda \mu^+\mu^-$ observables.
We would also like to thank Michal Kreps for useful discussion on the $\Lambda_b\to\Lambda \mu^+\mu^-$ branching fraction. 
DvD~is grateful to the Institute for Nuclear Theory at the University of Washington for their hospitality during which
a substantial part of this work was completed.\\

TB is supported by the Royal Society (United Kingdom).
SM is supported by the U.S. Department of Energy, Office of Science, Office of High Energy Physics under Award Number D{E-S}{C0}009913.
DvD~is supported by the Deutsche Forschungsgemeinschaft (DFG) within the Emmy Noether Programme under grant DY130/1-1
and the DFG Collaborative Research Center 110 ``Symmetries and the Emergence of Structure in QCD''.

\appendix

\bibliography{references}

\end{document}